\documentclass[prl,twocolumn,floatfix,superscriptaddress]{revtex4}
\usepackage{pstricks,graphicx,amsmath,bbm,mathrsfs,amssymb,psfrag,pifont,times}

\begin{document}

\title{Homodyne tomography with homodyne-like detection}

\author{Stefano~Olivares}
\affiliation{Quantum Technology Lab, Department of Physics ``Aldo Pontremoli'', University of Milan, via Celoria 16, I-20133 Milano (MI), Italy}
\affiliation{INFN, Sezione di Milano, via Celoria 16, I-20133 Milano (MI), Italy}
\author{Alessia~Allevi}\email{alessia.allevi@uninsubria.it}
\affiliation{Department of Science and High Technology, University of Insubria, Via Valleggio 11, I-22100 Como (CO), Italy}
\affiliation{Institute for Photonics and Nanotechnologies, CNR, Via Valleggio 11, I-22100 Como (CO), Italy}
\author{Giovanni~Caiazzo}
\affiliation{Department of Science and High Technology, University of Insubria, Via Valleggio 11, I-22100 Como (CO), Italy}
\author{Matteo~G.~A.~Paris}
\affiliation{Quantum Technology Lab, Department of Physics ``Aldo Pontremoli'', University of Milan, via Celoria 16, I-20133 Milano (MI), Italy}
\affiliation{INFN, Sezione di Milano, via Celoria 16, I-20133 Milano (MI), Italy}
\author{Maria~Bondani}
\affiliation{Institute for Photonics and Nanotechnologies, CNR, Via Valleggio 11, I-22100 Como (CO), Italy}

\begin{abstract}
We show that data from homodyne-like detection based on photon-number-resolving 
(PNR) detectors may be effectively exploited to reconstruct quantum states of 
light using the tomographic reconstruction techniques originally developed for 
homodyne detection based on photodiodes. Our results open new 
perspectives to quantum state reconstruction, and pave the way to the use of PNR-based 
homodyne-like detectors in quantum information science.
\end{abstract}

\maketitle

\par
\begin{figure}[tb]
\centering
\includegraphics[width=\columnwidth]{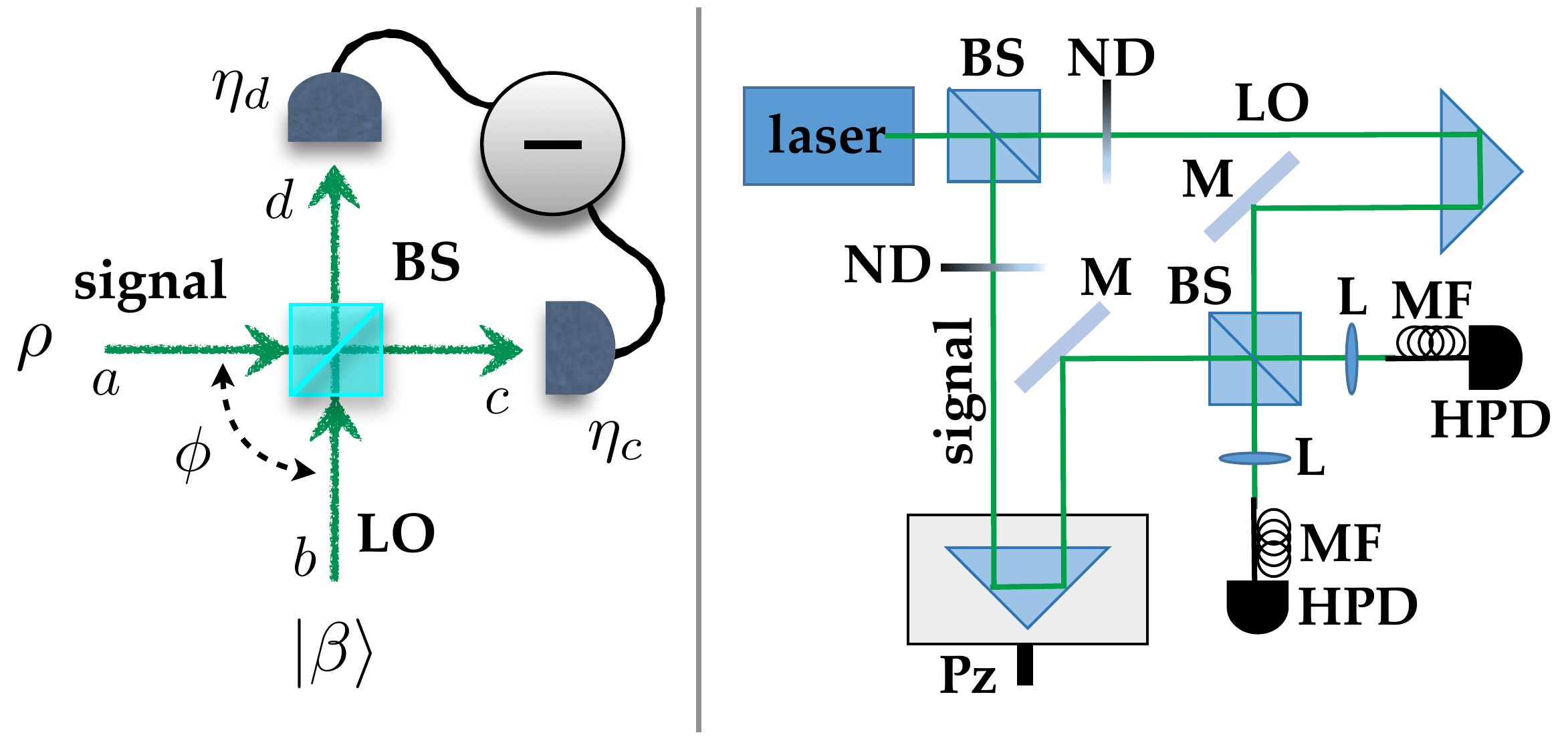}
\vspace{-.7cm}
\caption{(Left) HL detection scheme: the signal $\rho$ is mixed with a 
coherent state $| \beta \rangle$ (LO) at a balanced BS. After the 
interference, the difference between the number of photons detected 
at the BS outputs is evaluated.
(Right) Sketch of the experimental setup for the reconstruction of 
coherent and PHAV states.}
\label{setup}
\end{figure}

A quantum tomographic scheme is a technique to evaluate the expectation value 
of any observables and, in turn, to reconstruct the density matrix or the Wigner 
function of the system, by processing the outcomes of a {\it quorum of observables}, 
measured on repeated preparations of the state under investigation
\cite{gentomo,teo,paris,Dariano03}. 
In the continuous-variable domain, quantum states are usually reconstructed 
by means of optical homodyne tomography (OHT) \cite{Lvovsky08}. From the technical point of view, OHT is based on an interferometric scheme (see the left panel of Fig.~\ref{setup}) in which a state $\rho$ (the signal) is mixed at a balanced beam splitter (BS) with a high-intensity coherent state $| \beta \rangle = ||\beta| \, e^{i\phi}\rangle$ (the local oscillator, LO). The two outputs of the interferometer are detected by two pin photodiodes, whose difference photocurrent is suitably amplified, rescaled by the LO amplitude $|\beta|$, and recorded as a function of the LO phase $\phi$. By properly processing the data and applying advanced reconstruction algorithms, it is possible to achieve the complete knowledge of the state under investigation in terms of the density matrix \cite{Leo96,Dariano98} and to calculate the expectation values of given observables \cite{Dariano03}. Among them, we mention the inversion algorithms, which can be applied if the effective quantum efficiency of the detection apparatus is above 50~$\%$, and the maximum-likelihood reconstruction protocols, which can be applied also in the presence of strong losses \cite{hradil,rehacek,Puentes09,Strec09}.
\par
In this Letter we consider the homodyne-like (HL) detection scheme shown 
in the left panel of Fig.~\ref{setup}. At variance with the standard 
homodyne detection scheme, in our apparatus the pin {photodiodes} are 
replaced by photon-number-resolving detectors and the LO $|\beta \rangle$ 
is a low-intensity (few tens of photons) coherent state. The physical 
information is obtained from the difference between the effective 
number of detected photons, rather than from the difference between two 
macroscopic photocurrents. Recently, we have demonstrated that HL 
detection may be successfully exploited for state-discrimination of 
coherent states \cite{OE17,SciRep16}. Moreover, one has additional 
degrees of freedom available with HL data, and this may be used to 
outperform standard homodyne detection in quantum key distribution 
with continuous variables \cite{Cattaneo18}.
\par
Motivated by the results achieved so far, here we address quantum tomography 
of CV systems by the HL detection scheme \cite{IJQI17}. We prove that the HL 
scheme may be efficiently employed to reconstruct the density matrix of
single-mode quantum states, and to evaluate the expectation values of 
the first moments of the quadrature operator. In particular, we show that
robust and reliable results may be obtained with a relatively modest 
unbalancing between signal and LO. We also investigate the role played 
by quantum efficiency, which is a crucial parameter in the reconstruction 
of nonclassical states.
\par
In OHT, the elements of the density matrix $\rho$ of a quantum state can 
be reconstructed as follows \cite{gentomo,teo,paris,Dariano03,Lvovsky08,Dariano98,Leo96}
\begin{equation}\label{rhonm}
\rho_{nm} = \int_0^{\pi}\! d\phi \int_{-\infty}^{+\infty}\! 
dx\, p(x, \phi) F_{nm}(x, \phi) ,
\end{equation}
where $p(x, \phi) = \langle x_\phi | \varrho | x_\phi \rangle$ are the 
homodyne probability distributions and $| x_\phi \rangle$ are the eigenstates 
of the quadratures $\hat x_\phi = (\hat{a}\, e^{-i\phi} + \hat{a}^\dag\, e^{-i\phi})/\sqrt{2}$,
$\hat{a}$ being the annihilation operator of the field, $[\hat{a},\hat{a}^\dag] = 1$. In Eq.~(\ref{rhonm}), $F_{nm}(x, \phi)$ are a set of suitable sampling
functions (see Ref.~{\cite{Leo96}} for details). In practice,
the probabilities $p(x, \phi)$ are retrieved from the difference of 
the two macroscopic photocurrents exiting the homodyne detector \cite{lvovsky,cialdi16,esposito14}, and the matrix elements are obtained
by sampling $F_{nm}(x, \phi)$ over data.
\par
In our scheme, we replace $p(x, \phi)$ with the HL probability distributions $p_{\text{HL}}(\Delta_\phi)$, where $\Delta_\phi = \Delta/(\sqrt{2}|\beta|)$ and $\Delta$ is the (phase-dependent) difference between the number of photons detected at the two outputs of the homodyne detector \cite{OE17}. We notice that, while the outcome $x$ of standard homodyne detection can assume any real value, the quantity $\Delta_\phi$ is intrinsically discrete. However, this discretization does not represent a limitation for the proper reconstruction of the states under examination, as shown hereafter. The information coming from the HL scheme is contained in 
the joint statistics $q(n,m)$ of the number of photons detected 
at the outputs of the BS. To find the theoretical expression of $q(n,m)$, 
it is useful to consider the Glauber-Sudarshan $P$ representation 
of the input state $\rho$
\begin{equation}
\rho = \int_{\mathbb C} d^2 z\, P(z)\,  | z \rangle \langle z |\,,
\end{equation}
where $P(z)$ is the $P$-function associated with $\rho$.
After the interference of the input state with the LO (the coherent state $| \beta \rangle$) at the balanced BS, the two-mode input state
$R_{\text{in}} = \rho\otimes | \beta \rangle \langle \beta |$ becomes
\begin{equation}\label{R:output}
R_{\text{out}} = \int_{\mathbb C} d^2 z\, P(z)\,
\Big | \frac{\beta+z}{\sqrt{2}} \Big \rangle \Big \langle \frac{\beta+z}{\sqrt{2}}  \Big | \otimes
\Big | \frac{\beta-z}{\sqrt{2}} \Big \rangle \Big \langle \frac{\beta-z}{\sqrt{2}}  \Big |\,
\end{equation}
and the corresponding joint photon-number statistics can be written as \cite{braunstein,freyberger}:
\begin{align}
&q(n,m) =\nonumber\\
&\int_{\mathbb C}\!\! d^2 z\, P(z)\,
e^{-\mu_c(z,\beta) - \mu_d(z,\beta)}
\frac{\left[ \mu_c(z,\beta) \right]^n \left[ \mu_d(z,\beta) \right]^m}{n! m!}
\label{output:joint}
\end{align}
where $ \mu_c(z,\beta) = \frac12 |\beta + z|^2$ and $\mu_d(z,\beta) = \frac12 |\beta - z|^2$. We note that, thanks to this formalism, we can easily add the effect of quantum efficiency $\eta_k$ of the detector on the output modes $k = c,d$, by replacing $\mu_k(z,\beta)$ with $\eta_k \mu_k(z,\beta)$. Starting from the joint probability (\ref{output:joint}) it is possible to calculate the theoretical probability distribution of the quantity $\Delta = n-m$, namely
\begin{equation}\label{PHL}
p_{\text{HL}}(\Delta) = \left\{
\begin{array}{l l}
\sum_{k=0}^{\infty} q(\Delta+k,k) & \mbox{if}~\Delta \ge 0\,,\\
\sum_{k=1}^{\infty} q(k,k-\Delta) & \mbox{if}~\Delta < 0\,,
\end{array}
\right.
\end{equation}
and the corresponding ``rescaled'' version $p_{\text{HL}}(\Delta_\phi)$, where $\Delta_\phi = \Delta/(\sqrt{2}|\beta|)$.
In order to investigate the performance of the HL detector, we consider some paradigmatic optical states: the coherent state, which is phase-sensitive, its phase averaged counterpart (PHAV), which is of course phase insensitive, and the single-photon Fock state.
While for the coherent and the PHAV states we provide an experimental demonstration, for the Fock state we performed Monte Carlo simulated
experiments and we studied the effect of a non-unit quantum efficiency $\eta < 1$ with respect to the ideal case, namely $\eta = 1$.
\par
For a coherent state $\rho = | \alpha \rangle \langle \alpha |$, the 
theoretical joint photon-number statistics, and the HL distribution, 
can be obtained by setting $P(z) = \delta^{(2)}(z-\alpha)$, 
where $\delta^{(2)}(\zeta)$ is the complex Dirac's delta function. 
The experimental setup is sketched in the right panel of Fig.~\ref{setup}.
The second harmonic pulses ($\sim$ 5 ps pulse duration) at 523 nm of a mode-locked Nd:YLF laser regeneratively amplified at 500 Hz were divided at a Mach-Zehnder interferometer into two parts in order to yield the signal and the LO. The length of one arm of the interferometer was changed in steps by means of a piezoelectric movement (Pz) in order to vary the relative phase $\phi$ between the two arms in the interval $[0,\pi]$. Two variable neutral density filter wheels (ND) were used to change the balancing between the two beams, which were then recombined in a balanced BS. At the BS outputs two multi-mode 600-$\mu$m-core fibers (MF) delivered the light to a pair of photon-number-resolving detectors. In detail, we employed two hybrid photodetectors (HPDs, mod. R10467U-40, Hamamatsu Photonics), whose outputs were amplified, synchronously integrated and digitized. HPDs are commercial detectors endowed with partial photon-number-resolving capability and a good linearity up to 100 photons.  As already demonstrated elsewhere (see for instance \cite{JMO, arimondo, OE17}), HPD response can be characterized in a self-consistent way with the same light under examination. Here we just remark that from the experimental data it is possible to calculate the gain of the detection apparatus in order {to} recover the distribution of detected photons at each BS output. In addition, this kind of detector allows us to obtain information on the relative phase between the two arms of the interferometer by simply monitoring the mean number of detected photons measured at each BS output as a function of the piezoelectric movement \cite{JosaBwigner,OE17}. In the present case, $3 \times 10^6$ data corresponding to subsequent laser pulses were recorded and used to reconstruct the density matrix of the coherent state and of the PHAV state, which is a diagonal state obtained by randomizing the phase of a coherent state $|\alpha \rangle=   | |\alpha| \, e^{i\theta}\rangle$
\begin{equation} \label{rhophav}
\rho_{\rm PHAV} = \int_0^{2 \pi} \frac{d \theta}{2 \pi} |{\alpha} \rangle \langle {\alpha}| = e^{-|{\alpha}|^2} \sum_{n=0}^{\infty} \frac{|{\alpha}|^{2n}}{n!} |n \rangle \langle n|.
\end{equation}
\par
In order to process the data corresponding to the coherent state, once obtained the number of detected photons at each BS output, we calculated the shot-by-shot photon-number difference, $\Delta$. The corresponding phase value $\phi$ was determined by acquiring a suitable data sample ($5\times 10^4$ pulses) for each one of the 60 piezo positions. As the piezo moves in regular steps, the measured mean number of detected photons follows a sinusoidal trend due to the interference at the beam splitter, from which the effective value of $\phi$ for each piezo position can be evaluated. The $5\times 10^4$ data corresponding to the same $\phi$ were uniformly distributed around that value with a step of $1/(5\times 10^4)$. 
\begin{figure}[h!]
\includegraphics[width=\columnwidth]{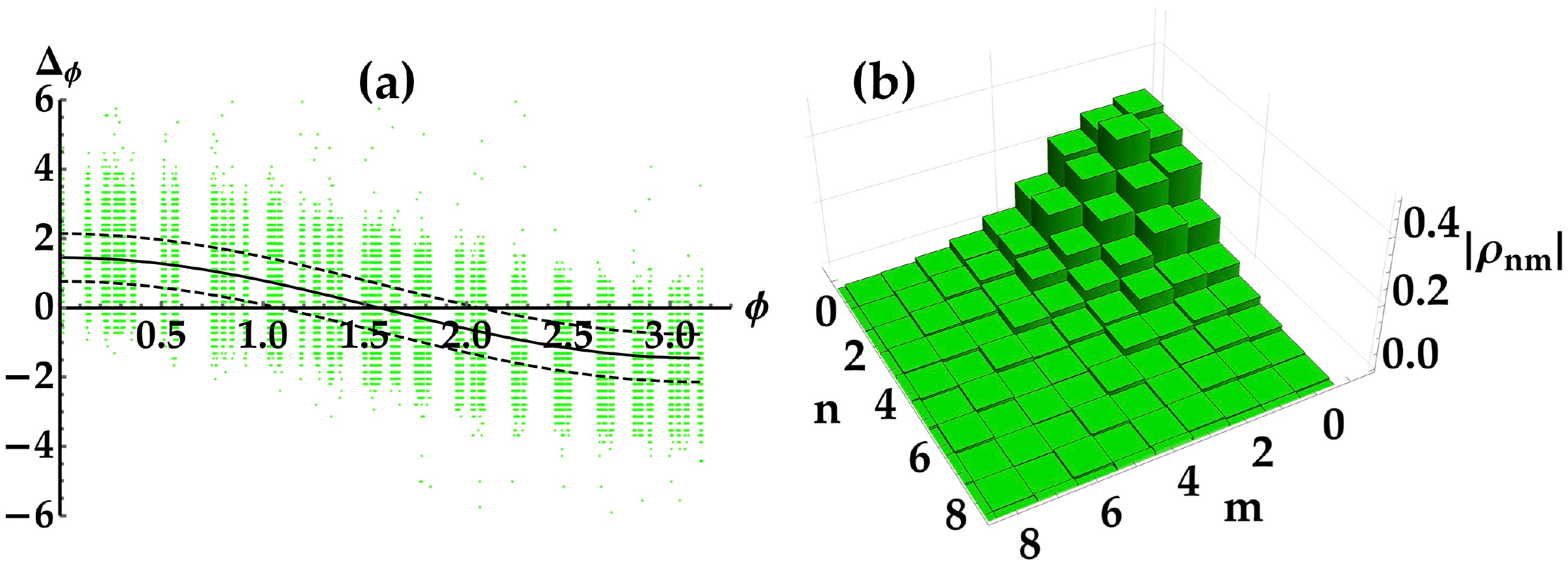}
\caption{Experimental reconstruction of a coherent state with 
measured amplitude $\alpha = 1.03$. The LO has an amplitude $|\beta| = 
3.82$. (a)  $\Delta_\phi$ as a function of LO phase, $\phi$. 
The experimental data (green dots) are shown together 
with the mean value (solid line) and
the standard deviation (dashed lines) of the quadrature, as 
obtained through the pattern function tomography. (b) 
Reconstructed density matrix. The fidelity 
with the expected state is $F = 99.4~\%$.}
\label{coherexp}
\end{figure}
\par
The typical behavior of $\Delta_\phi$ as a function of $\phi$ is shown in Fig.~\ref{coherexp}(a) ($3\times 10^5$ data). Note that, due to irregularities in 
the piezoelectric movement, the data are not uniformly distributed in $\phi$. 
The reconstruction procedure was applied to $10$ sets of $3 \times 10^5$ 
data drawn from the whole sample: the modulus of the average density matrix 
is shown in Fig.~\ref{coherexp}(b). The corresponding mean number of
 photons is $\langle n \rangle = \sum_n n\, \rho_{n,n} = 1.06$. The fidelity \cite{RJ:fidelity} between {the} reconstructed coherent state and a 
 coherent state with the same amplitude is $F = 99.4~\%$.
\par
As a matter of fact, the fidelity may not fully assess the quality of a
reconstruction scheme \cite{bina:fid} and therefore, in order to test
the quality of HL reconstruction, we also consider the HL evaluation of 
the expectation values  of some observables. To this aim, we remind that 
given a generic observable, we may obtain
its expectation value $\langle \hat{A} \rangle = \mbox{Tr}[\rho\hat{A}]$ 
from the distribution of homodyne data as \cite{Dariano03}
\begin{equation}\label{kernel}
\langle \hat{A} \rangle = \frac{1}{\pi}\int_0^{\pi} d\phi \int_{\mathbb{R}} dx \, p(x,\phi) \mathcal{R}[\hat{A}](x,\phi)
\end{equation}
where $\mathcal{R}[\hat{A}](x,\phi)$ is a kernel, or pattern function, associated with the operator $\hat{A}$. For instance, in the following we will use the fist 
two moments of the quadrature operator, which
may be obtained from the kernels \cite{Dariano03}:
\begin{align}
\mathcal{R}[\hat{x}_{\theta}](x,\phi) &=  2x \cos({\phi - \theta})\,, \\
\mathcal{R}[\hat{x}_{\theta}^2](x,\phi) &= \left(4x^2-\frac{1}{\eta}\right)\left[4 \cos^2 (\phi - \theta)-1\right]\frac{1}{4} + \frac{1}{4}\,.
\end{align}
\par
\begin{figure}[tb]
\centering
\includegraphics[width=\linewidth]{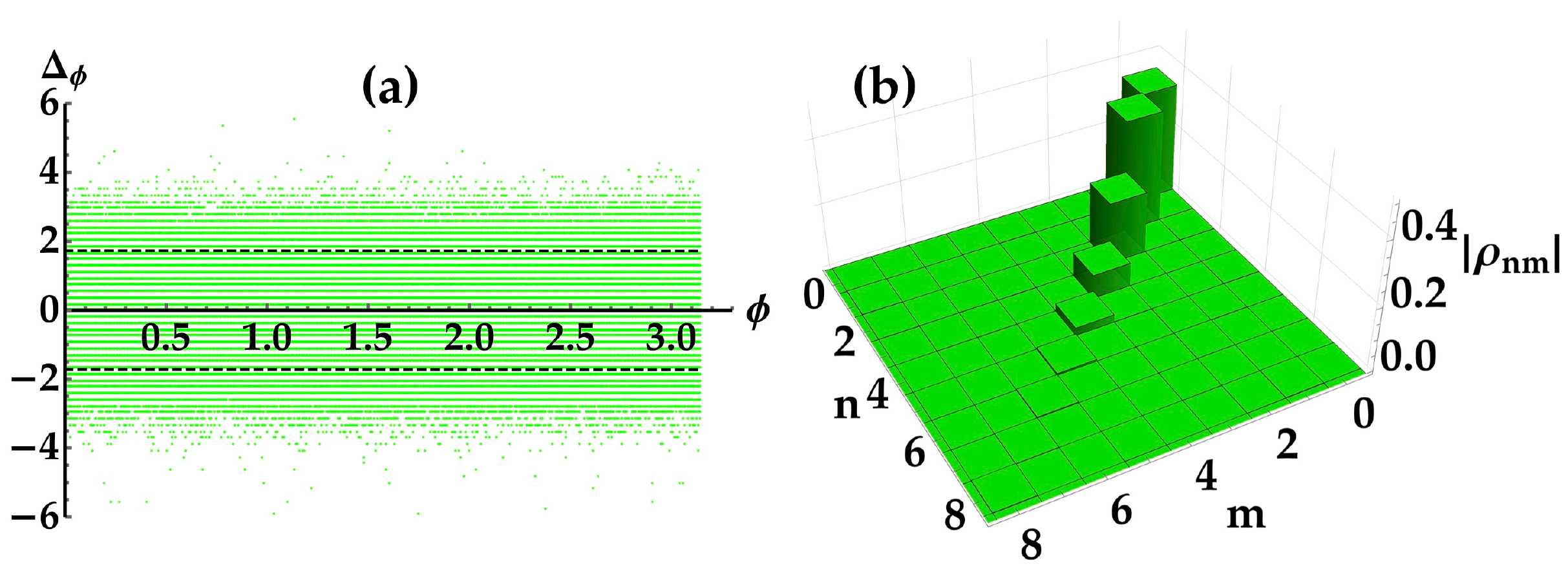}
\caption{Experimental reconstruction of a PHAV state with amplitude $|\alpha| = 1.08$. The LO has an amplitude $|\beta| = 3.82$. (a) $\Delta_{\phi}$ as a function of $\phi$. The experimental data (green dots) are shown together with the mean value (solid line) and
the standard deviation (dashed lines) of the quadrature as obtained through the pattern function tomography. (b) Reconstructed density matrix. 
The fidelity with the expected state is $F = 99.9~\%$.
}
\label{phavexp}
\end{figure}

By using the HL probabilities $p_{\rm HL}(\Delta_{\phi})$ instead of $p(x,\phi)$ and setting $\theta = 0$  
from  Eq.~(\ref{kernel}) we got: $\langle \hat{x}_{0} \rangle = 1.451 \pm 0.003$ and
$\mbox{var}[\hat{x}_0]=\langle \hat{x}^2_{0}\rangle - \langle\hat{x}_{0}\rangle^2 = 0.688 \pm 0.015$ for the case under investigation. These values must be compared to those expected in the limit of a HL scheme, namely $\langle \hat{x}_{0} \rangle_{\rm theo} = \sqrt{2}\, \alpha = \sqrt{2}~1.03 = 1.456$ and $\mbox{var}[\hat{x}_0]_{\rm theo} = \frac12 + \frac12 |\alpha|^2/|\beta|^2 = 0.536$. The last result explicitly shows the contribution to the variance due to the low-intensity LO, which becomes negligible as $|\beta| \gg |\alpha|$. Notice that since we are dealing with 
classical states, the quantum efficiency just rescales the energy of the detected states, and we can safely set $\eta=1$: the results we obtain are thus referred 
to the \emph{detected} states.  We see that while the mean value of the quadrature is well reconstructed by the experimental data, the variance is larger than expected. 
Such discrepancy is likely due to phase fluctuations occurred during the 
measurements session. In order to check whether this interpretation holds, we 
consider the PHAV state, which is phase insensitive and thus should not be 
influenced by the presence of possible phase fluctuations.
As in the case of the coherent state, we saved $10$ sets of $3\times 10^5$ data by calculating the shot-by-shot photon-number difference between the two BS outputs. Since the PHAV state is phase-insensitive, we randomly assigned a phase value to each experimental value of $d$. A typical trace of $d$ $vs.$ $\phi$ is shown in Fig.~\ref{phavexp}(a), whereas in panel (b) the modulus of the  reconstructed density matrix is presented. As expected, the off-diagonal elements are absent. In order to compare the reconstructed matrix to the theoretical prediction in Eq.~(\ref{rhophav}) with $|\alpha| = 1.08$, we calculated  the fidelity and obtained $F = 99.9~\%$. For what concerns the first and second moment of the quadrature, we obtained $\langle\hat{x}_{\theta}\rangle = 0.004 \pm 0.003$ and $\mbox{var}[\hat{x}_{\theta}] = 1.725 \pm 0.013$, $\forall \theta$. In this case, the expected values are $\langle\hat{x}_{\theta}\rangle_{\rm theo} = 0$ and $\mbox{var}[\hat{x}_{\phi}]_{\rm theo} = \frac12 +  |\alpha|^2 + \frac12 |\alpha|^2/|\beta|^2 = 1.706$, respectively. The very good agreement between theory and experiment for this phase-insensitive state confirms our considerations about the variance of the reconstructed coherent state.
\par
We now turn the attention to the Fock state $| 1 \rangle$, for which we show the results obtained by using Monte Carlo simulated experiments setting $\beta = \sqrt{20}$.
The state $| 1 \rangle$ has a highly singular $P$ function given by:
\begin{equation}
P(z) = \delta^{(2)}(z) + \frac{\partial^2}{\partial z\, \partial z^*} \delta^{(2)}(z)\,,
\end{equation}
and the joint probability in Eq.~(\ref{output:joint}) reads:
\begin{align}
q_\eta &(n,m) = \nonumber\\
&\frac{e^{-\eta |\beta|^2}}{n!\, m!} \left(\frac{\eta |\beta|^2}{2}\right)^{n+m}
\left[
1 +\frac{(n-m)^2 - \eta |\beta|^2}{|\beta|^2}
\right]\,,
\end{align}
where we assumed that both the detectors have the same quantum efficiency $\eta$. The corresponding $p_{\text{HL}}(\Delta)$ is quite clumsy and is not explicitly reported here. 
\begin{figure}[h!]
\centering
\includegraphics[width=\columnwidth]{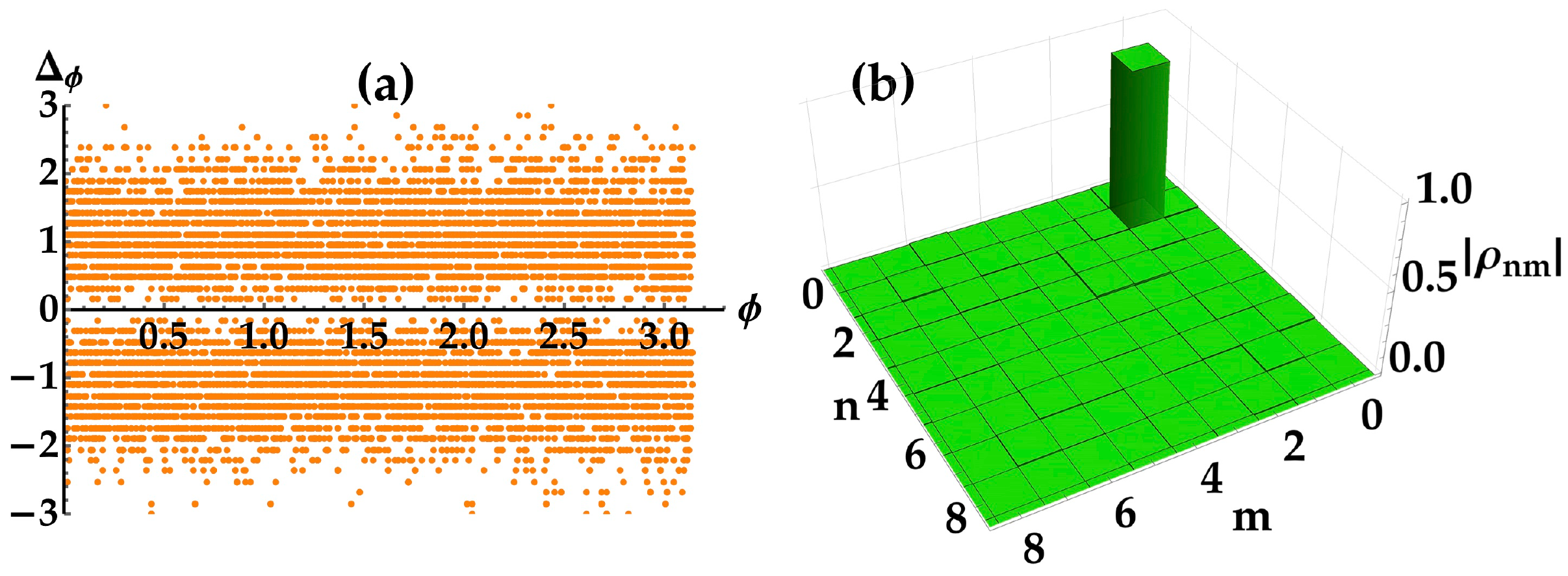}
\caption{Reconstruction of the Fock state $| 1 \rangle$ (Monte Carlo simulated experiment), with LO amplitude $\beta = \sqrt{20}$ and quantum efficiency $\eta = 1$. (a) $\Delta_\phi$ as a function of $\phi$; (b) Reconstructed density matrix. The fidelity with the expected state is $F = 99.9~\%$.
}
\label{focketa1}
\end{figure}
\par
In Fig.~\ref{focketa1}(a), we show a typical HL trace, {\it i.e.} $\Delta_\phi$ as a function of $\phi$, obtained from a Monte Carlo simulation
(with $5 \times 10^4$ data) assuming ideal detection, {\it i.e.} $\eta = 1$. It is clear that, as one may expect, $\Delta_\phi$ does not depend on the relative phase $\phi$ since the Fock state is phase insensitive. The modulus of the reconstructed density matrix is plotted in panel Fig.~\ref{focketa1}(b). In this case the density matrix is diagonal and thus the fidelity of the state coincides with that of the photon-number distribution. In particular, for the density matrix in Fig.~\ref{focketa1}(b) we have $F = 99.9~\%$.
\par
\begin{figure}[h!]
\includegraphics[width=\linewidth]{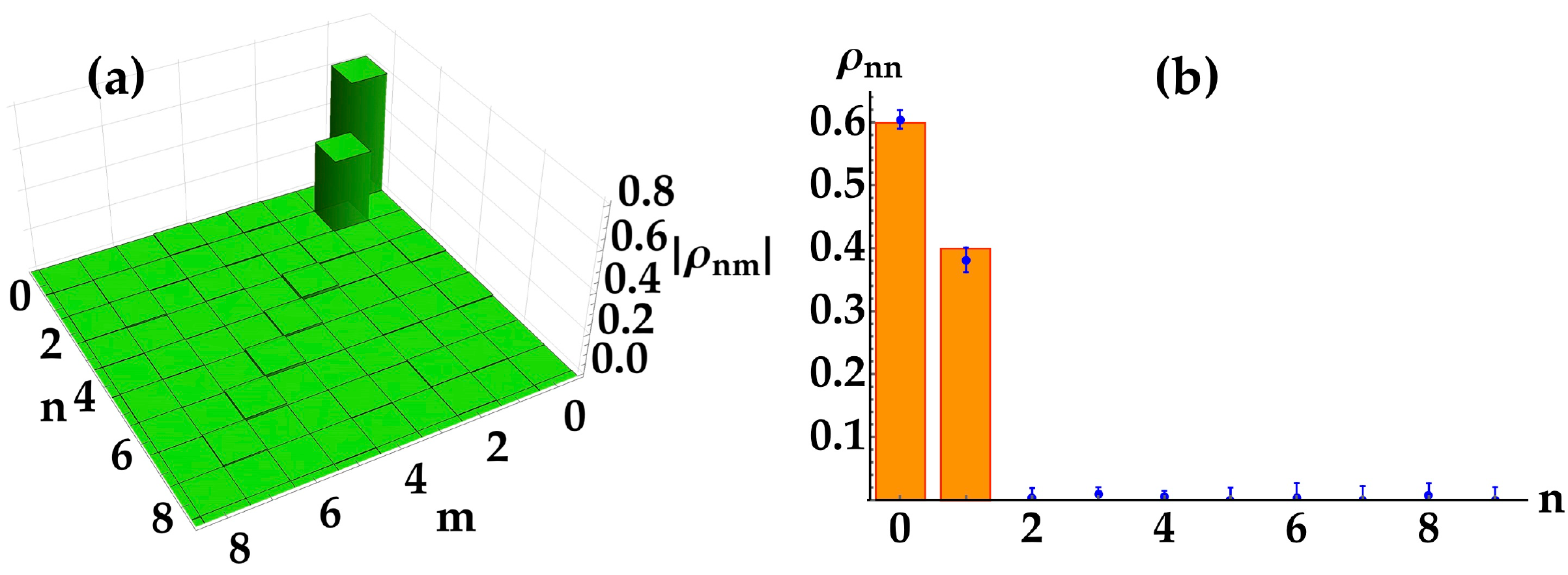}
\caption{Reconstruction of the Fock state $| 1 \rangle$ (Monte Carlo simulated experiment), with LO amplitude $\beta = \sqrt{20}$ and
quantum efficiency $\eta = 0.4$. (a) {Reconstructed} density matrix; (b) Photon-number statistics of the reconstructed state.
(the error bars have been obtained averaging over 10 simulated experiments). The fidelity with respect to the expected distribution is $F = 99.0~\%$.
}
\label{simuletaless1}
\end{figure}

Up to this point, we have shown the reliability of the HL detection scheme 
and of the reconstruction strategy under ideal detection conditions. However, 
since photon-number-resolving detectors are real detectors, it is worth 
investigating whether the HL scheme also works in the presence of an 
overall quantum efficiency $\eta <1$. When standard HD is adopted, it 
is well known that a non-unit quantum efficiency rescales the mean 
value of the reconstructed density matrix in the case of classical 
states of light, such as coherent and PHAV states, whereas it deeply 
modifies the properties of the reconstruction in the case of 
nonclassical states, such as Fock states \cite{lvovsky}.
In order to check if analogous results can be achieved by means 
of the HL scheme, we present the results obtained by simulated experiments 
for the Fock state $| 1 \rangle$. We assume the same value of LO 
adopted above, {\it i.e.}  $\beta = \sqrt{20}$, but now we set $\eta = 
0.4$, a realistic value for many kinds of photon-number-resolving detectors \cite{JMO,SiPMarxiv}. 
In Fig.~\ref{simuletaless1}, we plot the modulus of the reconstructed 
density matrix, panel (a), and {the photon-number} statistics, panel (b), for the Fock state $| 1\rangle$. It is clear from the reconstructed density matrix that the effect of a {non-unit} quantum efficiency is to add a vacuum component to the state:
in this case the expected density matrix is $\rho = \eta |1 \rangle \langle 1 | + (1-\eta) | 0 \rangle \langle 0 |$ \cite{lvovsky} and its fidelity with respect {to} the reconstructed one is $F = 99.0~\%$.
\par
In conclusion, our numerical and experimental results demonstrate 
that the HL detection scheme may be used, instead of standard (pin-based) 
homodyne detectors, to reconstruct the density matrix of quantum states of light, 
as well as the first two moments of the quadrature operator.
Our results open new perspectives 
to quantum state reconstruction, and pave the way to the use of 
PNR-based homodyne-like detectors in quantum information science.


\begin{thebibliography}{99}
\bibitem{gentomo} G.~M.~D'Ariano, L.~Maccone, M.~G.~A.`Paris,
{\em Quorom of observables for universal quantum estimation},
J. Phys. A {\bf 34}, 93 (2001).
\bibitem{teo} Y. S. Teo, \emph{Introduction to Quantum-state Estimation}, (World Scientific, Singapore, 2016).
\bibitem{paris} M.~G.~A. Paris, J. $\check{\rm R}$eh$\acute{\rm a}\check{\rm c}$ek (Eds.), \emph{Quantum-state Estimation,} Lect. Not. Phys. (Springer, 2004).
\bibitem{Dariano03} G.~M.~D’Ariano, M.~G.~A.~Paris, and M.~F.~Sacchi, \emph{Quantum Tomography,} Adv. Imaging Electron. Phys. \textbf{128}, 205--308 (2003).
\bibitem{Lvovsky08} A.~I.~Lvovsky, and M.~G.~Raymer, \emph{Continuous-variable optical quantum state tomography,}, Rev. Mod. Phys. \textbf{81}, 299-332 (2009).
\bibitem{Dariano98} G.~M.~D'Ariano, M.~Vasilyev, and P.~Kumar, \emph{Self-homodyne tomography of a twin-beam state,} Phys. Rev. A \textbf{58}, 636--648 (1998).
\bibitem{Leo96} U.~Leonhardt, M.~Munroe, T.~Kiss, Th.~Richter, and M.~G.~Raymer, \emph{Sampling of photon statistics and density matrix using homodyne detection,} Opt. Commun. \textbf{127}, 144-160 (1996). 
\bibitem{hradil} Z.~Hradil, \emph{Quantum-state estimation,} Phys. Rev. A \textbf{55}, 1561-1564(R) (1997).
\bibitem{rehacek} J.~$\check{\rm R}$eh$\acute{\rm a}\check{\rm c}$ek, Z.~Hradil, E.~Knill, and A.~I.~Lvovsky, \emph{Diluted maximum-likelihood algorithm for quantum tomography}, Phys. Rev. A \textbf{75}, 042108 (2007).
\bibitem{Puentes09} G.~Puentes, J.~S.~Lundeen, M.~P.~A. Branderhorst, H.~B. Coldenstrodt-Ronge, B.~ J.~Smith, and I.~A.~Walmsley, \emph{Bridging particle and wave sensitivity in a configurable detector of positive operator-valued measures,} Phys. Rev. Lett. \textbf{102}, 080404 (2009).
\bibitem{Strec09} A.~Allevi, A.~Andreoni, M.~Bondani, G.~Brida, M.~Genovese, M.~Gramegna, P.~Traina, S.~Olivares, M.~G.~A. Paris, and G.~Zambra, \emph{State reconstruction by on/off measurements,} Phys. Rev. A \textbf{80}, 022114 (2009).
\bibitem{SciRep16} M.~Bina, A.~Allevi, M.~Bondani, and S.~Olivares, \emph{Phase-reference monitoring in coherent-state discrimination assisted by a photon-number resolving detector,} Sci. Rep. \textbf{6}, 26025 (2016).
\bibitem{OE17} M.~Bina, A.~Allevi, M.~Bondani, and S.~Olivares, \emph{Homodyne-like detection for coherent state-discrimination in the presence of phase noise,} Opt. Express \textbf{25}, 10685-10692 (2017).
\bibitem{Cattaneo18} M.~Cattaneo, S.~Olivares, and M.~G.~A.~Paris, \emph{Hybrid quantum key distribution using coherent states and photon-number-resolving detectors,} Phys. Rev. A \textbf{98}, 012333 (2018).
\bibitem{IJQI17} A.~Allevi, M.~Bina, S.~Olivares, and M.~Bondani, \emph{Homodyne-like detection scheme based on photon-number-resolving detectors,} Int. J. Quantum Inf. \textbf{15}, 1740016 (2017).
\bibitem{cialdi16} S.~Cialdi, C.~Porto, D.~Cipriani, S.~Olivares, and M.~G.~A.~Paris, \emph{ Full quantum state reconstruction of symmetric two-mode squeezed thermal states via spectral homodyne detection and a state-balancing detector,} Phys. Rev. A \textbf{93}, 043805 (2016).
\bibitem{lvovsky} A.~I.~Lvovsky, H.~Hansen, T.~Aichele, O.~Benson, J.~Mlynek, and S.~Schiller, \emph{State Reconstruction of the Single-Photon Fock State,} Phys. Rev. Lett. \textbf{87}, 050402 (2001).
\bibitem{esposito14}  M.~Esposito, F.~Benatti, R.~Floreanini, S.~Olivares, F.~Randi, K.~Titimbo, M.~Pividori, F.~Novelli, F.~Cilento, F.~Parmigiani, and D.~Fausti, \emph{Pulsed homodyne Gaussian quantum tomography with low detection efficiency,} New. J. Phys. \textbf{16}, 043004 (2014).
\bibitem{braunstein} S.~L. Braunstein, \emph{Homodyne statistics,} Phys. Rev. A \textbf{42}, 474-481 (1990).
\bibitem{freyberger} M. Freyberger, K. Vogel and W. Schleich, \emph{From photon counts to quantum phase,} Phys. Lett. A \textbf{176}, 41-46 (1993).
\bibitem{JMO} M.~Bondani, A.~Allevi, A.~Agliati, and A.~Andreoni, \emph{Self-consistent characterization of light statistics,} J. Mod. Opt. \textbf{56}, 226-231 (2009).
\bibitem{arimondo} A.~Allevi, and M.~Bondani, \emph{Nonlinear and quantum optical properties and applications of intense twin-beams,} Adv. At. Mol. Opt. Phys. \textbf{66}, 49-110 (2017).
\bibitem{JosaBwigner} M.~Bondani, A.~Allevi, and A.~Andreoni, \emph{Self-consistent phase determination for Wigner function reconstruction,} J. Opt. Soc. Am. B \textbf{27}, 333-337 (2010).
\bibitem{RJ:fidelity} R.~Jozsa, \emph{Fidelity for Mixed Quantum States,} J. Mod. Opt. \textbf{41}, 2315–2323 (1994).
\bibitem{bina:fid} M. Bina, A. Mandarino, S. Olivares, and M. G. A. Paris, \emph{Drawbacks of the use of fidelity to assess quantum resources,} Phys. Rev. A \textbf{89}, 012305 (2014).
\bibitem{SiPMarxiv} G.~Chesi, L.~Malinverno, A.~Allevi, R.~Santoro, A.~Martemiyanov, M.~Caccia, and M.~Bondani, \emph{Optimizing Silicon photomultipliers for Quantum Optics}, to be submitted.
\end{thebibliography}
\end{document}